\documentclass[prc,preprint,showpacs,showkeys,
tightenlines,
        amsfonts,nofootinbib]{revtex4}

\usepackage{graphicx}  %
\usepackage{bm}  %
\usepackage{amssymb}
\usepackage{dcolumn}

\begin{document}

%


\newcommand{\beq}{\begin{equation}}
\newcommand{\eeq}{\end{equation}}
\newcommand{\bea}{\begin{eqnarray}}
\newcommand{\eea}{\end{eqnarray}}
\newcommand{\ben}{\begin{eqnarray*}}
\newcommand{\een}{\end{eqnarray*}}

\newcommand{\simlt}{\stackrel{<}{{}_\sim}}
\newcommand{\simgt}{\stackrel{>}{{}_\sim}}
\newcommand{\sing}{$^1\!S_0$ }
\newcommand{\btau}{\mbox{\boldmath$\tau$}}
\newcommand{\bsig}{\mbox{\boldmath$\sigma$}}

\newcommand{\dt}{\partial_t}

\newcommand{\kf}{k_{\rm F}}
\newcommand{\wt}{\widetilde}
\newcommand{\kt}{\widetilde k}
\newcommand{\pt}{\widetilde p}
\newcommand{\qt}{\widetilde q}
\newcommand{\wh}{\widehat}
\newcommand{\dens}{\rho}
\newcommand{\edens}{{\cal E}}
\newcommand{\order}[1]{{\cal O}(#1)}

\newcommand{\psihat}{\widehat\psi}
\newcommand{\dagphan}{{\phantom{\dagger}}}
\newcommand{\kvec}{{\bf k}}
\newcommand{\kpvec}{{\bf k}'}
\newcommand{\ak}{a^\dagphan_\kvec}
\newcommand{\akdag}{a^\dagger_\kvec}
\newcommand{\akv}[1]{a^\dagphan_{\kvec_{#1}}}
\newcommand{\akdagv}[1]{a^\dagger_{\kvec_{#1}}}
\newcommand{\akp}{a^\dagphan_{\kvec'}}
\newcommand{\akpdag}{a^\dagger_{\kvec'}}
\newcommand{\akpv}[1]{a^\dagphan_{\kvec'_{#1}}}
\newcommand{\akpdagv}[1]{a^\dagger_{\kvec'_{#1}}}

\def\vec#1{{\bf #1}}

\newcommand{\nab}{\overrightarrow{\nabla}}
\newcommand{\nabsq}{\overrightarrow{\nabla}^{2}\!}
\newcommand{\nabl}{\overleftarrow{\nabla}}
\newcommand{\galnab}{\tensor{\nabla}}
\newcommand{\psid}{{\psi^\dagger}}
\newcommand{\psidal}{{\psi^\dagger_\alpha}}
\newcommand{\psidbe}{{\psi^\dagger_\beta}}
\newcommand{\idt}{{i\partial_t}}
\newcommand{\Sthree}{{\delta_{11'}(\delta_{22'}\delta_{33'}%
        -\delta_{23'}\delta_{32'})%
        +\delta_{12'}(\delta_{23'}\delta_{31'}-\delta_{21'}\delta_{33'})%
        +\delta_{13'}(\delta_{21'}\delta_{32'}-\delta_{22'}\delta_{31'})}}
\newcommand{\Stwo}{{\delta_{11'}\delta_{22'}-\delta_{12'}\delta_{21'}}}
\newcommand{\Left}{{\cal L}}
\newcommand{\Tr}{{\rm Tr}}

\newcommand{\h}{\hfil}
\newcommand{\be}{\begin{enumerate}}
\newcommand{\ee}{\end{enumerate}}
\newcommand{\I}{\item}   

\newcommand{\density}{\rho}

\newcommand{\thyp}{\mbox{---}}

\newcommand{\Jks}{J_{\ks}}
\newcommand{\Jzero}{\Jks}
\newcommand{\Jdensityzero}{J_\density^0}

\newcommand{\ks}{{\rm ks}}
\newcommand{\Seq}{Schr\"odinger\ equation }
\newcommand{\yvec}{{\bf y}}
\newcommand{\ve}{V_{eff}}
\newcommand{\densityJzero}{\density^0_J}
\newcommand{\Dfunct}{{D^{-1}}}
\newcommand{\drv}[2]{{\mbox{$\partial$} #1\over \mbox{$\partial$} #2}}
\newcommand{\drvs}[2]{{\mbox{$\partial^2$} #1\over \mbox{$\partial$} #2 \mbox{$^2$}}}
\newcommand{\drvt}[2]{{\partial^3 #1\over \partial #2 ^3}}
\newcommand{\til}[1]{{\widetilde #1}}
\newcommand{\dthreex}{d^3\xvec}
\newcommand{\dthreey}{d^3\yvec}

\newcommand{\efermi}{\varepsilon_{{\scriptscriptstyle \rm F}}}
\newcommand{\eHF}{\wt\varepsilon}
\newcommand{\eKS}{e}
\newcommand{\ekJ}{e_\kvec^J}
\newcommand{\epsk}{\varepsilon_\kvec}
\newcommand{\epsKS}{\varepsilon}
\newcommand{\Eq}[1]{Eq.~(\ref{#1})}

\newcommand{\Fi}[1]{\mbox{$F_{#1}$}}
\newcommand{\fq}{f_{\qvec}}

\newcommand{\Gammaalt}{\overline\Gamma}
\newcommand{\Gammaks}{\Gamma_{\ks}}
\newcommand{\Gamint}{\widetilde{\Gamma}_{\rm int}}
\newcommand{\GKS}{G_{\ks}}
\newcommand{\grad}{{\bm{\nabla}}}   
\newcommand{\greenKS}{{G}_{\ks}}
\newcommand{\greenKSp}{{G}_{\ks'}}
\newcommand{\intint}{\int\!\!\int}

\newcommand{\kfermi}{k_{{\scriptscriptstyle \rm F}}}   
\newcommand{\kJzero}{k_J}

\renewcommand{\l}{\lambda}

\newcommand{\MeV}{\mbox{\,MeV}}
\newcommand{\mi}[1]{\mbox{$\mu_{#1}$}}

\newcommand{\Oi}[1]{\mbox{$\Omega_{#1}$}}

\newcommand{\phibar}{\overline\phi}
\newcommand{\phidagger}{\phi^\dagger}
\newcommand{\phistar}{\phi^\ast}
\newcommand{\psibar}{\overline\psi}
\newcommand{\psidagger}{\psi^\dagger}
\newcommand{\qvec}{\vector{\rho}}

\newcommand{\tr}{{\rm tr\,}}

\newcommand{\Ulong}{U_{L}}

\renewcommand{\vector}[1]{{\bf #1}}
\newcommand{\vext}{v_{\rm ext}}   
\newcommand{\Vlong}{V_{L}}

\newcommand{\Wzero}{W_0}
\newcommand{\Wks}{W_{\ks}}
\newcommand{\xvec}{\vector{\bf{x}}}
\newcommand{\zvec}{\vector{z}}
\newcommand{\rvec}{\vector{r}}

\newcommand{\ts}{\textstyle}
%
%
%
%


%
\title{Single-Particle Properties from Kohn--Sham Green's Functions}
\author{Anirban Bhattacharyya}\email{anirban@mps.ohio-state.edu}
\author{R.J. Furnstahl}\email{furnstahl.1@osu.edu}

\affiliation{Department of Physics,
         The Ohio State University, Columbus, OH\ 43210}

\date{\today}
\date{October, 2004}

\begin{abstract}
%
An effective action approach to Kohn--Sham density
functional theory is used to illustrate how the exact Green's
function can be calculated in terms of the
Kohn--Sham Green's function.
An example based on Skyrme energy functionals shows that
single-particle Kohn--Sham spectra can be improved by
adding sources used to construct the energy functional. 
\end{abstract}

\smallskip
\maketitle


The Skyrme-Hartree--Fock approach to nuclear properties has had wide
success in reproducing bulk properties of nuclei across the periodic
table
\cite{VB72,RINGSCHUCK,BROWN98,Dobaczewski:2001ed,BENDER2003,Stoitsov:2003pd}.  
The interpretation of
the Skyrme formalism as an approximate implementation of Kohn--Sham
density functional theory (DFT) \cite{BRACK85} implies that
certain observables (energy per particle, densities) can be
calculated reliably, but
these do not include single-particle quantities.
Only for the bulk observables can the DFT framework accommodate all
correlations in principle (if not in practice because of the limited
form of the energy functionals actually used)
\cite{KOHN65,PARR89,DREIZLER90}.
Nevertheless, single-particle energies and wave functions
from Skyrme and other DFT-like formalisms are also regularly
used.
 
In this letter, 
we illustrate how to extend the effective action approach to
Kohn--Sham DFT \cite{VALIEV96,RASAMNY98,PUG02,FURNSTAHL04b} 
to calculate the full single-particle Green's function in
terms of Kohn--Sham Green's functions at the same level of
approximation.
Our discussion directly adapts  
the extension described in the context of Coulomb systems
in Refs.~\cite{VALIEV97,VALIEV97b}.
This connection between Green's functions
helps to clarify both some misconceptions and
limitations of the Kohn--Sham approach, and suggests how to improve
calculations of single-particle properties.
At first,
we consider functionals of the fermion density only, and then compare
to generalized functionals that also depend on the kinetic energy density
to illustrate the effect of additional sources.


We introduce a generating functional in the path integral formulation
with a Lagrangian ${\cal L}$ 
supplemented by a local c-number
source $J(x)$ coupled 
to the composite density operator as in Ref.~\cite{PUG02},
but add a non-local c-number
source $\xi(x,x')_{\alpha\beta}$ coupled to $\psi_\alpha(x)\psi_\beta^\dag(x')$,   
\beq
    Z[J,\xi] = e^{iW[J,\xi]}
      = \int\! D\psi D\psi^\dag \ e^{i\int\! d^4x\ [{\cal L}\,+\,   J(x)
    \psi_\alpha^\dag(x) \psi_\alpha(x)\,+\,
    \int\! d^4x'\, \psi_\alpha(x)\xi(x,x')_{\alpha\beta}\psi_\beta^\dag(x')]}
    \ ,
   \label{partfunc1}
\eeq
where $\alpha$ and $\beta$ are spin indices and summation of repeated
indices is implied.
(We generalize below to an additional local source, as in
Ref.~\cite{FURNSTAHL04b}.)
For simplicity, normalization factors are considered to be implicit
in the functional integration measure \cite{FUKUDA94,FUKUDA95}.
As a specific example, we will use the effective field theory (EFT)
Lagrangian appropriate for a dilute Fermi system \cite{HAMMER00},
but the discussion can be adapted to any system for which a hierarchy of
approximations can be defined.

The fermion density in the presence of the sources $J$ and $\xi$ is
\beq
  \rho(x)\equiv \langle \psi_\alpha^\dag(x)\psi_\alpha (x)\rangle_{J\,,\,\xi}
   = \frac{1}{iZ}  \frac{\delta Z[J,\xi]}{\delta J(x)}
   = \frac{\delta W[J,\xi]}{\delta J(x)} \ .
\label{partden1}
\eeq
Note that the sources here are time dependent, in contrast to the more
limited discussion with static sources in Ref.~\cite{PUG02}; 
however, the generalization of the formalism is direct.
A functional Legendre transformation from $J$ to $\rho$, which takes us
from $W$ to the effective action $\Gamma$, produces an
energy functional of the density, which is minimized at the exact 
ground-state density for time-independent sources.%
\footnote{Note that 
the energy functional is only obtained once $\xi$ is set to zero.}
The inversion method \cite{FUKUDA94,FUKUDA95} 
carries out this inversion order-by-order in a
specified expansion; an EFT expansion was used in Refs.~\cite{PUG02} and
\cite{FURNSTAHL04b}.
At the end, one sets $J(x)$ and $\xi(x,x')$ to zero.
(Although we are unaware of any general problems,
we have not excluded the possibility of 
complications in making the inversions with time-dependent sources.)

Solving the zeroth-order system 
defines the Green's
function $\greenKS(x,x')_{\alpha\beta}$ 
of the Kohn--Sham non-interacting system
in the presence of $\xi(x,x')_{\alpha\beta}$, the Kohn--Sham potential
$J_0(x)$, and an external potential $v({\bf x})$.
This Green's function satisfies 
\beq
  \int\! d^4z\ [\greenKS(x,z)]_{\alpha\gamma}^{-1}
  \greenKS(z,x')_{\gamma\beta}
  = \delta_{\alpha\beta}\delta^4(x-x') \ ,
\eeq
or  
\beq  
  \int\! d^4z \left[\left( i \partial_t+ 
    \frac{\bm{\nabla}^2}{2M} 
    -v(\xvec)+J_0(x)
  \right)\delta_{\alpha\gamma}\delta^4(x-z) - \xi(z,x)_{\gamma\alpha} \right] 
  \greenKS (z,x')_{\gamma\beta} = \delta_{\alpha\beta}\delta^4(x-x')
 \ ,
\label{eq:GKSEq}
\eeq
with appropriate finite-density boundary conditions (one could also
introduce a chemical potential).
Note that $\greenKS$ doesn't take a  simple form in terms of orbitals 
[see $\greenKS^0$ in Eq.~(\ref{eq:Gks})] until we set 
$\xi = 0$ and restrict ourselves to time independent $J_0$. 

Functional derivatives of $W$ with respect to $\xi(x,x')$
gives the two-point
function in the presence of the sources,  
\beq
  iG(x,x')_{\alpha\beta} \equiv
    \langle T [\psi_\alpha(x)\psi_\beta^\dag(x')] \rangle_{J,\xi}
   = \frac{1}{iZ}  \frac{\delta Z[J,\xi]}{\delta\xi(x,x')_{\alpha\beta}}
   = \frac{\delta W[J,\xi]}{\delta\xi(x,x')_{\alpha\beta}}
  \ .  
  \label{eq:defexactG}
\eeq
The exact ground-state Green's function is obtained by
setting $\xi = J= 0$ 
after taking derivatives.
The key results we will need to evaluate Eq.~(\ref{eq:defexactG}) 
in terms of Kohn--Sham quantities were given in
Refs.~\cite{VALIEV97,VALIEV97b} (we follow their notation for the most
part) and are rederived here.
First, functional derivatives with respect to $\xi$ of $W$ and
$\Gamma$ are directly related, 
where
\beq
  \Gamma[\rho,\xi] = W[J,\xi] - \int\! d^4y\, J(y)\rho(y)
\eeq
is the effective action.
Namely, the functional derivative with respect to $\xi$
of this equation yields (spin indices are suppressed)
\beq
  \left(
    \frac{\delta \Gamma[\rho,\xi]}{\delta\xi(x,x')}
  \right)_{\rho}
 =
  \left(
    \frac{\delta W[J,\xi]}{\delta\xi(x,x')}
  \right)_{J}
  +
  \int\!d^4 y
  \left(
    \frac{\delta W[J,\xi]}{\delta J(y)}
  \right)_{\xi}
  \left(
    \frac{\delta J(y)}{\delta\xi(x,x')}
  \right)_{\rho} 
  -
  \int\!d^4 y
  \left(
    \frac{\delta J(y)}{\delta\xi(x,x')}
  \right)_{\rho} \, \rho(y)
  \ ,
\eeq
from which the last two terms cancel, leaving
\beq
  \left(
    \frac{\delta W[J,\xi]}{\delta\xi(x,x')_{\alpha\beta}}
  \right)_{J}
  =
  \left(
    \frac{\delta \Gamma[\rho,\xi]}{\delta\xi(x,x')_{\alpha\beta}}
  \right)_{\rho}
  \ .
  \label{eq:same}
\eeq
(Here and below we repeatedly apply the functional relations
\beq
  \left( \frac{\delta F}{\delta\xi} \right)_{\rho}
  =
  \left( \frac{\delta F}{\delta\xi} \right)_{J}
  +
  \left( \frac{\delta F}{\delta J} \right)_{\xi}
  \left( \frac{\delta J}{\delta\xi} \right)_{\rho}
  =
  \left( \frac{\delta F}{\delta\xi} \right)_{J}
  -
  \left( \frac{\delta F}{\delta\rho} \right)_{\xi}
  \left( \frac{\delta \rho}{\delta\xi} \right)_{J}
  \ ,
  \label{eq:relations}
\eeq
where $F=F[J,\xi]$ and arguments and integrals are implied.)
Equation~(\ref{eq:same}) is a special case of a general
result for Legendre transformations proved in Ref.~\cite{ZINNJUSTIN}.

Next,  this relation applied to the zeroth-order
(Kohn--Sham) system yields the Kohn--Sham Green's function,
\beq
  \left(
    \frac{\delta \Gamma_0[\rho,\xi]}{\delta\xi(x,x')_{\alpha\beta}}
  \right)_{\rho}
  =
  \left(
    \frac{\delta W_0[J_0,\xi]}{\delta\xi(x,x')_{\alpha\beta}}
  \right)_{J_0}
  =
  i \greenKS(x,x')_{\alpha\beta}
  \ .
  \label{eq:same0}
\eeq
We divide the full effective action into zeroth-order and interacting pieces,
\beq
  \Gamma[\rho,\xi] = \Gamma_0[\rho,\xi] + \Gamma_{\rm int}[\rho,\xi]
  \ .
\eeq
Since $\Gamma_{\rm int}[\rho,\xi]$ depends on $\xi$ only through
$\greenKS$, 
\beq
  \left(
    \frac{\delta \Gamma_{\rm int}[\rho,\xi]}{\delta\xi(x,x')_{\alpha\beta}}
  \right)_{\rho}
  = 
  \int\!\!\int\! 
  \frac{\delta \Gamma_{\rm int}[\rho,\xi]}
    {\delta\greenKS(y,y')_{\delta\gamma}}\,
  \left(
    \frac{\delta \greenKS(y,y')_{\delta\gamma}}{\delta\xi(x,x')_{\alpha\beta}}
  \right)_{\rho}
  \ dy\,dy'
  \ .
  \label{eq:Gammaint}
\eeq
The second half of the integrand can be rewritten
\bea
  \left(
    \frac{\delta \greenKS(y,y')_{\delta\gamma}}{\delta\xi(x,x')_{\alpha\beta}}
  \right)_{\rho}
  & = &
  \left(
    \frac{\delta
    \greenKS(y,y')_{\delta\gamma}}{\delta\xi(x,x')_{\alpha\beta}}
  \right)_{J_0}
  +
  \int\! 
  \left(
    \frac{\delta \greenKS(y,y')_{\delta\gamma}}{\delta J_0(z)}
  \right)_{\xi}
    \left(
    \frac{\delta J_0(z)}{\delta\xi(x,x')_{\alpha\beta}}
  \right)_{\rho}
   \, dz  
  \nonumber
  \\ 
  & = &
  \left(
    \frac{\delta \greenKS(y,y')_{\delta\gamma}}{\delta\xi(x,x')_{\alpha\beta}}
  \right)_{J_0}
  -
  \int\! 
  \left(
    \frac{\delta \greenKS(y,y')_{\delta\gamma}}{\delta \rho(z)}
  \right)_{\xi}
    \left(
    \frac{\delta \rho(z)}{\delta\xi(x,x')_{\alpha\beta}}
  \right)_{J_0}
   \, dz  
  \nonumber
  \\ & = &
  \greenKS(x,y')_{\alpha\gamma}\greenKS(y,x')_{\delta\beta}
    \nonumber
   \\ & & \qquad \null
    + i \int\! 
  \left(
    \frac{\delta \greenKS(y,y')_{\delta\gamma}}{\delta\rho(z)}
  \right)_{\xi}
  \greenKS(x,z)_{\alpha\lambda}\greenKS(z,x')_{\lambda\beta}\, dz
  \ .
  \label{eq:secondhalf}
\eea
The second line follows by applying Eq.~(\ref{eq:relations}) with
$F \rightarrow \greenKS$ and simplifying.
The functional derivatives in the second line 
can be evaluated by using the expression for $\greenKS$ in
terms of the noninteracting  generating functionals.
Thus,
\bea
  \left( \frac{\delta\greenKS(y,y')_{\delta\gamma}}
              {\delta\xi(x,x')_{\alpha\beta}} 
  \right)_{J_0} 
  &=&
    \frac{\delta}{\delta\xi(x,x')_{\alpha\beta}}
    \left[
    -\frac{1}{Z_0} \frac{\delta Z_0[J_0,\xi]}{\delta\xi(y,y')_{\delta\gamma}}
    \right]
   \nonumber \\
  &=&
   \left[
     \frac{1}{Z_0} 
     \frac{\delta Z_0[J_0,\xi]}
          {\delta\xi(x,x')_{\alpha\beta}}
   \right]
   \left[
     \frac{1}{Z_0} 
     \frac{\delta Z_0[J_0,\xi]}
          {\delta\xi(y,y')_{\delta\gamma}}
   \right]
   - \frac{1}{Z_0} 
     \frac{\delta^2 Z_0[J_0,\xi]}
          {\delta\xi(x,x')_{\alpha\beta}\,\delta\xi(y,y')_{\delta\gamma}}
   \nonumber \\
  &=&
   i^2 \langle
     T [\psi_\alpha(x) \psidagger_\beta(x')]
          \rangle \,
          \langle
     T [\psi_\delta(y) \psidagger_\gamma(y')]
          \rangle
    - i^2
      \langle
     T [\psi_\alpha(x) \psidagger_\beta(x')
       \psi_\delta(y) \psidagger_\gamma(y')]
          \rangle
   \nonumber \\
  &=&
   (-i)^2 \langle
     T [\psi_\alpha(x) \psidagger_\gamma(y')]
          \rangle \,
          \langle
     T [\psi_\delta(y) \psidagger_\beta(x')]
          \rangle
   \nonumber \\
  &=&
   \greenKS(x,y')_{\alpha\gamma} \,
   \greenKS(y,x')_{\delta\beta}
   \ ,
\eea
where we've applied Wick's theorem to the noninteracting system
to go from the third line to the fourth line,
and
\beq
  \left( \frac{\delta\rho(z)}
              {\delta\xi(x,x')_{\alpha\beta}} 
  \right)_{J_0}   
  =
   -i \left( \frac{\delta\greenKS(z,z^+)_{\delta\delta}}
              {\delta\xi(x,x')_{\alpha\beta}} 
      \right)_{J_0} 
  =  
   -i \greenKS(x,z)_{\alpha\delta} \,
      \greenKS(z,x')_{\delta\beta}
      \ .
\eeq
Alternatively, we can expand $\delta (\greenKS\greenKS^{-1})/\delta \xi =
0$ and use $\delta \greenKS^{-1}/\delta\xi = -1$.

Substituting Eq.~(\ref{eq:secondhalf}) 
back into Eq.~(\ref{eq:Gammaint}), we find that 
\bea
  \left(
    \frac{\delta \Gamma_{\rm int}[\rho,\xi]}{\delta\xi(x,x')_{\alpha\beta}}
  \right)_{\rho}
  & = &
  \int\!\!\int\! 
  \greenKS(x,y')_{\alpha\gamma}\, 
  \frac{\delta \Gamma_{\rm
  int}[\rho,\xi]}{\delta\greenKS(y,y')_{\delta\gamma}}\,
  \greenKS(y,x')_{\delta\beta}\ dy\,dy'
  \nonumber \\ &  &
  \null + i \int\! \greenKS(x,y)_{\alpha\lambda}
  \left(\frac{\delta \Gamma_{\rm
  int}[\rho,\xi]}{\delta\rho(y)}\right)_\xi
  \greenKS(y,x')_{\lambda\beta}\ dy
  \ .
  \label{eq:complicated}
\eea
Equations~(\ref{eq:same0}) and (\ref{eq:complicated}), together with
\beq
  i G(x,x')_{\alpha\beta}
    = \left(
       \frac{\delta W[J,\xi]}{\delta\xi(x,x')_{\alpha\beta}}
      \right)_{J}
    = \left(
       \frac{\delta \Gamma[\rho,\xi]}{\delta\xi(x,x')_{\alpha\beta}}
      \right)_{\rho}
    = \left(
       \frac{\delta \Gamma_{0}[\rho,\xi]}{\delta\xi(x,x')_{\alpha\beta}}
      \right)_{\rho}
    + \left(
       \frac{\delta \Gamma_{\rm int}[\rho,\xi]}{\delta\xi(x,x')_{\alpha\beta}}
      \right)_{\rho}
  \ ,
\eeq
imply a Dyson equation for the exact Green's
function: 
\beq
  G(x,x')_{\alpha\beta} = \greenKS(x,x')_{\alpha\beta} + 
    \int\!\greenKS(x,y')_{\alpha\gamma}\, 
       \Sigma_{\rm ks}(y',y)_{\gamma\delta}\, 
       \greenKS(y,x')_{\delta\beta}
    \ dy\,dy'
  \ ,
  \label{eq:exactG}
\eeq
which defines a self-energy $\Sigma_{\rm ks}$ as
\bea
  \Sigma_{\rm ks}(y',y)_{\gamma\delta} & = &
  \frac{1}{i} \frac{\delta \Gamma_{\rm int}[\rho,\xi]}
       {\delta\greenKS(y,y')_{\delta\gamma}}\,
  + \left(
      \frac{\delta \Gamma_{\rm int}[\rho,\xi]}{\delta\rho(y)}
    \right)_\xi
  \delta_{\gamma\delta}\,\delta^4(y'-y)
  \nonumber \\
  & \equiv &
  \Sigma'_{\rm ks}(y',y)_{\gamma\delta} 
    + J_0(y')\,\delta_{\gamma\delta}\,\delta^4(y'-y)
  \ .
  \label{eq:Sigma0b}
\eea
In the second line, the self-consistent Kohn--Sham potential $J_0$ is equal
to $\delta\Gamma_{\rm int}/\delta\rho$ only when we set $J=0$ \cite{PUG02}.
Neither $\Sigma_{\rm ks}$ nor $\Sigma'_{\rm ks}$ is the conventional
self-energy, which is built from non-interacting (rather than Kohn--Sham)
Green's functions.
We can obtain $\Sigma'_{\rm ks}(y',y)$ at the diagrammatic level by
opening each $\greenKS$ line in turn in a given Feynman diagram
for $\Gamma_{\rm int}$.
It consists of the same diagrams as the conventional
one-particle-reducible self-energy, but with the fermion lines given
by $\greenKS$ rather than the non-interacting Green's function (which
includes only the external potential).


%
\begin{figure}[t]
\centerline{\includegraphics*[width=4in,angle=0]{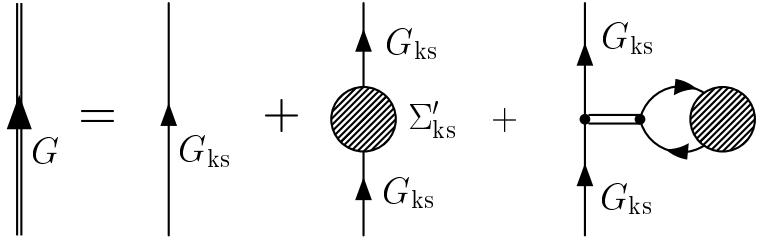}}
\vspace*{-.1in}
\caption{Equation for the full propagator in terms of the Kohn--Sham
Green's functions and self-energy.}
\label{fig:fullG}
\end{figure}        

Now consider applying these equations with $\xi = J = 0$ \emph{after}
taking functional derivatives; we denote the Kohn--Sham Green's function
in this case as $\greenKS^0$.
For simplicity we will consider spin-independent interactions, so that
the Green's functions and self-energies are diagonal in spin.
Kohn--Sham orbitals arise as solutions to
\beq
  \bigl[ 
  -\frac{\bm{\nabla}^2}{2M} 
  +  v(\xvec) -  J_0(\xvec)
  \bigr]\, \psi_k({\bf x}) = \varepsilon_k \psi_k({\bf x})
  \, ,
 \label{eq:ks1}  
\eeq 
where the index $k$ represents all quantum numbers except for the spin
\cite{PUG02}.
The decomposition of $\greenKS^0(x,x')_{\alpha\beta} =
\delta_{\alpha\beta} \greenKS^0(x,x')$ in terms
of these orbitals is \cite{PUG02} 
\beq
   i\greenKS^0 (\xvec t,\xvec't')=\sum_k \psi_k (\xvec)\,
       \psi_k^* (\xvec')\,  
   e^{-i\varepsilon_k(t-t')}[\theta(t-t')\, 
           \theta(\varepsilon_k-\varepsilon_{\rm F})
   -\theta(t'-t)\, \theta(\varepsilon_{\rm F}-\varepsilon_k)]
  \label{eq:Gks}
  \ ,
\eeq
corresponding (in frequency space) to simple poles, just like a Hartree
Green's function.
It is well known that the Kohn--Sham single-particle 
eigenvalues $\varepsilon_k$ are
not physical except at the Fermi surface \cite{PARR89,DREIZLER90}.
Nevertheless,
the trace of this Green's function gives the complete ground-state
density $\rho(\xvec)$ (that is, the exact result if we calculate to all
orders).

\begin{figure}[t]
\centerline{\includegraphics*[width=5.5in,angle=0]{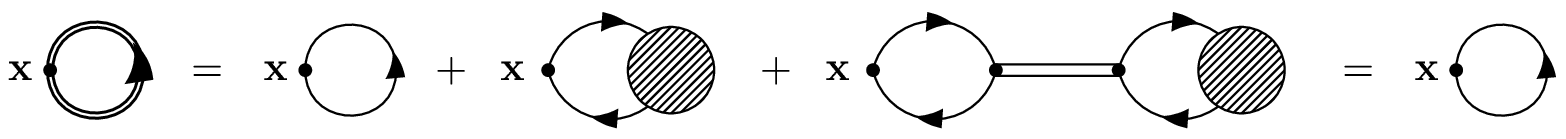}}
\vspace*{-.1in}
\caption{Equation for the density, showing the equivalence of the
full and Kohn--Sham densities.}
\label{fig:densities}
\end{figure}

\begin{figure}[t]
\centerline{\includegraphics*[width=3.5in,angle=0]{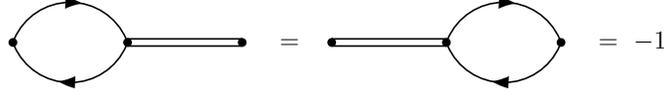}}
\vspace*{-.1in}
\caption{Cancellation of the density-density correlator with
$\delta J_0/\delta\rho$.}
\label{fig:cancellation}
\end{figure}

We can easily show diagrammatically
that Eq.~(\ref{eq:exactG}) implies that
the density obtained from the Kohn--Sham Green's function is, as
advertised, exactly equal to that obtained from the exact Green's
function. 
The density can be directly expressed in
terms of the Kohn--Sham Green's function with equal arguments as
\beq
  \rho(\xvec) = -i\,\greenKS^0(x,x^+)_{\alpha\alpha} 
  = -i\nu\,\greenKS^0(x,x^+)  \ ,
  \label{eq:denG} 
\eeq 
where $\nu$ is the spin-isospin degeneracy.
In Fig.~\ref{fig:fullG}, we have rewritten the last term in the
Dyson equation (\ref{eq:exactG}) for the exact Green's function using
\beq
  \frac{\delta\Gamma_{\rm int}}{\delta\rho({\bf y})}
  =
  \int\! 
  \frac{\delta\Gamma_{\rm int}}{\delta J_0({\bf z})}
  \frac{\delta J_0({\bf z})}{\delta\rho({\bf y})}
  \ d^3{\bf z}
  \ ,
\eeq
where  
$\delta J_0({\bf z})/\delta\rho({\bf y}) = [\delta^2
W_0/\delta J_0({\bf y})\delta J_0({\bf z})]^{-1}$,
which is minus the inverse density-density correlator
\cite{VALIEV97,PUG02}, 
is represented with a double
line (with no arrow).
The  result of carrying out Eq.~(\ref{eq:denG}) on Eq.~(\ref{eq:exactG})
is shown in Fig.~\ref{fig:densities}, where the last two diagrams cancel
as in Fig.~\ref{fig:cancellation}.
Note that while similar cancellations were shown in Ref.~\cite{PUG02} in the
special case of zero-range interactions, the result here is completely
general.
Thus we see that the exact density is reproduced by the Kohn--Sham
Green's function \emph{by construction}.


To illustrate some issues in comparing Kohn--Sham and exact
Green's functions,
we apply the formalism with the effective Lagrangian for dilute Fermi
systems used in prior investigations:
\bea
  {\cal L}  &=&
       \psi^\dagger \biggl[i\partial_t + \frac{\nab^{\,2}}{2M}\biggr]
                 \psi - \frac{C_0}{2}(\psi^\dagger \psi)^2
            + \frac{C_2}{16}\Bigl[ (\psi\psi)^\dagger
                                  (\psi\galnab^2\psi)+\mbox{ h.c.}
                             \Bigr]
  \nonumber \\[5pt]
   & & \null +
         \frac{C_2'}{8} (\psi \galnab \psi)^\dagger \cdot
             (\psi\galnab \psi)
+  \cdots \ ,
  \label{eq:lag}                                                   
\eea
where $\galnab=\overleftarrow{\nabla}-\nab$ is the Galilean invariant
derivative and h.c.\ denotes the Hermitian conjugate.
To describe trapped fermions,
we add to the Lagrangian a  
term for an external confining potential $v(\xvec)$  
coupled to the density 
operator $v(\xvec)\psi^\dagger\psi$ \cite{PUG02}.
For the numerical calculations, we take the potential 
to be an isotropic harmonic confining potential,
\beq
  v(\xvec)= \frac{1}{2} m\,\omega^2 \, |\xvec|^2 \ ,
 \label{eq:vext}
\eeq
although the discussion holds for a general non-vanishing
external potential.

We repeat the previous development
to introduce a second energy functional with an additional
local source coupled to the kinetic energy
density, following Ref.~\cite{FURNSTAHL04b}.
The comparison of results from the two functionals 
illustrates how the Kohn--Sham single-particle spectrum can be
significantly
different even though the bulk observables are essentially equal
\cite{FURNSTAHL04b}. 
So we consider
\beq
 Z'[J,\eta,\xi] = e^{iW'[J,\eta,\xi]}
      = \int\! D\psi D\psi^\dag \ e^{i\int\! d^4x\ [{\cal L}\,+\,   J
    \psi^\dag \psi\,+\,
    \eta\bm{\nabla}\psi^\dag\cdot\bm{\nabla}\psi
    \,+\,
    \int\! d^4x'\, \psi(x)\xi(x,x')\psi^\dag(x')]}
    \ ,
   \label{partfunc2}
\eeq
and the corresponding effective action
\beq
  \Gamma'[\rho,\tau,\xi] = W'[J,\eta,\xi] - \int\! d^4y\, J(y)\rho(y) 
     - \int\! d^4y\ \eta(y)\tau(y)
     \ , 
\eeq
with kinetic energy density
\beq
  \tau(x) \equiv 
    \langle \bm{\nabla}\psi^\dag(x)\bm{\cdot\nabla}\psi
    (x)\rangle_{J\,,\,\eta\,,\,\xi}
          = \frac{\delta W'[J,\eta,\xi]}{\delta \eta(x)} 
          \ .
  \label{partkin1}
\eeq
(We use superscript primes on the functionals, and $\mbox{ks}'$ on the
self-energies and Green's functions to distinguish the following
quantities from those without $\eta$ or $\tau$ dependence.) 
Each step goes through with straightforward generalizations,
yielding Eq.~(\ref{eq:exactG}) again, but now with
\bea
  \Sigma_{\rm ks'}(y',y)_{\gamma\delta} & = &
  \frac{\delta \Gamma'_{\rm
  int}[\rho,\tau,\xi]}{\delta\greenKSp(y,y')_{\delta\gamma}}\,
  + [J_0(y')
  + \bm{\nabla}_{y'}\bm{\cdot} \bm{\nabla}_{y} \,
    \eta_0(y')]\, \delta_{\gamma\delta}\, \delta^4(y'-y)
  \nonumber \\
  & \equiv &
  \Sigma'_{\rm ks'}(y',y)_{\gamma\delta}  + [J_0(y')
  + \bm{\nabla}_{y'}\bm{\cdot} \bm{\nabla}_{y} \,
    \eta_0(y')]\, \delta_{\gamma\delta}\, \delta^4(y'-y)
  \label{eq:Sigma0}
\eea
after $J(y')$ and $\eta(y')$ are set to zero.
[Note that the gradients act on the $\greenKSp$'s to produce
$\tau$ after partial
integrations in Eq.~(\ref{eq:exactG}).]

These two functionals were compared in Ref.~\cite{FURNSTAHL04b}
for a dilute gas of fermions in a harmonic trap.
Two sets of parameters were used to illustrate the impact of a larger
effective mass $M^*(\xvec)$, which appears only in the ``$\rho\tau$''
(primed) formalism.  
Even though the fermion density and energy per particle for
the $\rho$ and $\rho\tau$ functionals were very similar, 
the single-particle spectra have significant and systematic differences
(see Ref.~\cite{FURNSTAHL04b} for details and figures).
We can understand the systematics of the difference by comparing
Kohn--Sham and exact spectra for a uniform system.
We will drop the non-Hartree--Fock terms, which have been treated in LDA
in both cases and which contribute equally to the energy spectra. 
We note that 
the terms in the $\rho\tau$ functional 
correspond directly with terms in conventional Skyrme energy
functionals \cite{FURNSTAHL04b}.

In the $\rho$ case, the Kohn--Sham equation for the single-particle
orbital \cite{PUG02} (with external potential set to zero) leads to
the spectrum
\beq
  \varepsilon_k^{\rho} = \frac{k^2}{2M} - J^{\rho}_0 
  \ ,
\eeq
where
\beq
  J_0^{\rho} = - \frac{\nu-1}{\nu} \, C_0  \rho
   -\left(
    (\nu -1)\frac{C_2}{15\pi^2}
    + (\nu+1)\frac{C'_2}{15\pi^2}
   \right) \left( \frac{6\pi^2\rho}{\nu}\right)^{5/3}
  \ .
\eeq
In the $\rho\tau$ case, we find a different spectrum
\beq
  \varepsilon_k^{\rho\tau} = \frac{k^2}{2M^*} - J^{\rho\tau}_0 
  \ ,
\eeq
where
\beq
  J_0^{\rho\tau} = - \frac{\nu-1}{\nu}\,C_0\rho
  - \left(
  \frac{\nu-1}{\nu}\, \frac{C_2}{4}
    +  \frac{\nu+1}{\nu}\, \frac{C'_2}{4}
  \right) \tau
  \ ,
\eeq
and
\beq
  \frac{1}{2M^{*}}
  = \frac{1}{2M} - {\eta_0}
  = \frac{1}{2M} + 
  \left(
     \frac{\nu-1}{\nu}\, \frac{C_2}{4}
    +  \frac{\nu+1}{\nu}\, \frac{C'_2}{4}
  \right)
  \rho
  \ .
 \label{eq:ks1b}  
\eeq 
Using $\tau = \frac{3}{5} \kf^2 \rho$, the difference
in the spectra simplifies to
\beq
  \varepsilon^{\rho}_k - \varepsilon^{\rho\tau}_k
    = \left(
     \frac{\nu-1}{\nu}\, \frac{C_2}{4}
    +  \frac{\nu+1}{\nu}\, \frac{C'_2}{4}
    \right) (\kf^2 - k^2) \rho
    \ .
\eeq
Thus, the spectra differ for all momentum states except
at the Fermi surface, where the spectra coincide as expected in Kohn--Sham DFT.
In detail, the $\rho\tau$ spectrum includes explicit momentum dependence
that is converted to density dependence (i.e., $\kf$ dependence) in the
$\rho$ spectrum.
We can also compare the Kohn-Sham 
spectra to that of the Green's function in
the Hartree--Fock approximation, where 
we find that the $\varepsilon^{\rho\tau}$
spectrum is the same as the Hartree--Fock spectrum.  Indeed, for this
approximation the $J_0$ and $\eta_0$ terms in Eq.~(\ref{eq:Sigma0})
precisely cancel against $\Sigma'_{\rm ks'}$.
In contrast, Eq.~(\ref{eq:Sigma0b}) yields a net contribution that
shifts the Kohn--Sham spectrum to the Hartree--Fock spectrum.

This example illustrates how individual  
Hartree--Fock self-energies in a gradient expansion 
can be completely included by adding the 
corresponding source terms.
(A different example with covariant energy functionals is given
in Ref.~\cite{FURNSTAHL04c}.)
The exact cancellations are only possible for local self-energies,
which means Hartree--Fock.
Beyond Hartree--Fock, the single-particle spectrum from the Kohn--Sham and
exact Green's functions will necessarily differ.
We can anticipate that self-energies with large non-localities will
lead to the most significant differences.
This is consistent with the expectation that low-lying vibrational
states can account for the difference in level density between Skyrme
(or other mean-field) and experimental spectra near the Fermi surface
\cite{RINGSCHUCK,BENDER2003}.


In this work, we have illustrated the relationship between Kohn--Sham and
exact Green's functions within an effective action formalism.
This approach goes beyond the observation that single-particle
properties are not reliably calculated in terms of Kohn--Sham orbitals
and eigenvalues.
The formalism presents two ways to improve single-particle spectra.
The Kohn--Sham spectra became closer to the exact spectra with the
addition of appropriate sources.
It is tempting to conclude that adding additional sources can always
improve the Kohn--Sham single-particle spectrum, but this will require
tests beyond the Hartree--Fock level.
More generally, Eq.~(\ref{eq:exactG}) shows how to calculate single-particle
quantities in terms of Kohn--Sham propagators at the same level
of approximation (which is determined by the truncation of
$\Gamma_{\rm int}$).
In future work,
the formalism will be applied to the calculation of spectral functions
and the effect of low-lying vibrational states on the spectra tested 
by including
self-energy diagrams that sum particle-hole bubbles.

\acknowledgments

We thank M.~Birse, A.~Bulgac, H.-W.~Hammer, S.~Puglia,
A.~Schwenk, and B.~Serot for useful comments and discussions.
This work was supported in part by the National Science Foundation
under Grants No.~PHY--0098645 and No.~PHY--0354916.


\end{document}